\documentclass[9pt,twocolumn,twoside]{pnas-new}
\templatetype{pnasresearcharticle} % Choose template 
\title{%Controlled wave break in excitable media using sub-threshold perturbations.
A mechanism for electric turbulence in cardiac tissue with optogenetic modification.% in the stable parameter r\'egime.
}

\author[a,1]{Rupamanjari Majumder}
\author[a,b]{Sayedeh Hussaini} 
\author[a]{Vladimir S. Zykov}
\author[a,b]{Stefan Luther}
\author[a,b,c]{Eberhard Bodenschatz}
\affil[a]{Max Planck Institute for Dynamics and Self-Organization, G\"ottingen, Germany}
\affil[b]{Institute for Dynamics of Complex Systems, University of Göttingen, D-37073 G\"ottingen, Germany}
\affil[c]{Laboratory of Atomic and Solid-State Physics and Sibley School of Mechanical and Aerospace Engineering, Cornell University, Ithaca, New York 14853, USA}

\leadauthor{Majumder} 

\significancestatement{Wave breaks occur in many complex excitable systems and often lead to self-sustained re-entry. In the heart, such re-entry causes fatal rhythm disturbances, so-called arrhythmias, the prevention of which requires a detailed understanding of the initiation and dynamics of these breaks. Optogenetics provides a promising tool to investigate and control wave dynamics in cardiac tissue. Although it is very effective at stimulating light intensities, the tools’ performance remains largely unknown at non-stimulating intensity levels. Here we demonstrate a new mechanism of wave break initiation in cardiac tissue, which is triggered by periodic sub-threshold illumination. The mechanism involves "conditioning" the wavelength of a propagating wave, which leads to the creation of new excitation windows for re-entry.}

\authorcontributions{Please provide details of author contributions here.}
\authordeclaration{The authors declare no competing interests.}
\correspondingauthor{\textsuperscript{1}Rupamanjari Majumder. E-mail: rupamanjari.majumder\@ds.mpg.de}

\keywords{Wave break $|$ Spiral waves $|$ Optogenetics $|$ Sub-threshold illumination $|$ Mechanism} 

\begin{abstract}
Interruptions in nonlinear wave propagation, commonly referred to as wave breaks, are typical of many complex excitable systems. In the heart they lead to fatal rhythm disorders, the so-called arrhythmias, which are one of the main causes of sudden death in the industrialized world. Progress in the treatment and therapy of cardiac arrhythmias requires a detailed understanding of the triggers and dynamics of these wave breaks. In particular, two very important questions are: 1) What determines the potential of a wave break to initiate re-entry? and 2) How do these breaks evolve such that the system is able to maintain spatiotemporally chaotic electrical activity? Here we approach these questions numerically using optogenetics in an \textit{in silico} model of human atrial tissue that has undergone chronic atrial fibrillation (cAF) remodelling. In the lesser known sub-threshold illumination r\'egime, we discover a new mechanism of wave break initiation in cardiac tissue that occurs for gentle slopes of the restitution characteristics. This mechanism involves "conditioning" or reshaping the wave profile from front to back, such that, removal of the external light source causes rapid recovery of cells at the waveback, leading to the creation of vulnerable windows for sustained re-entry in spatially extended systems.
\end{abstract}

\dates{This manuscript was compiled on \today}
\doi{\url{www.pnas.org/cgi/doi/10.1073/pnas.XXXXXXXXXX}}

\begin{document}

\maketitle
\thispagestyle{firststyle}
\ifthenelse{\boolean{shortarticle}}{\ifthenelse{\boolean{singlecolumn}}{\abscontentformatted}{\abscontent}}{}

% If your first paragraph (i.e. with the \dropcap) contains a list environment (quote, quotation, theorem, definition, enumerate, itemize...), the line after the list may have some extra indentation. If this is the case, add \parshape=0 to the end of the list environment.
\dropcap{S}piral waves occur as short- or long-lived transients in various natural excitable systems~\cite{Palsson1996,Belousov1985,Zhabotinsky1991,Bar1994,Krinsky1991,Kapral1995}.
In the heart they appear as abnormal electrical waves which are the basis of fatal cardiac arrhythmias (tachycardia and fibrillation)~\cite{Davidenko1990}. Regardless of the type of medium that sustains these waves, spirals exhibit some common dynamic characteristics, such as drift~\cite{Panfilov7922}, meander~\cite{Otani1993, ZHANG1995661}, anchoring~\cite{Steinbock_Mueller1993,Kuklik2008, Boily2007,Olmos2010}, detachment~\cite{Kitahata2018,PORJAI2016283,Feng2014,Tanaka2009} and breakup~\cite{Fenton2002}. Spiral wave breakup or turbulence is of particular interest to cardiac researchers because it is associated with fibrillation, the most common precursor of stroke and sudden death. In particular, the mechanisms underlying the occurrence of spiral wave turbulence in heart tissue are extensively investigated because of their clinical significance and possible implications~\cite{Luther2011}. %An widely-used method for studying and removing electrical turbulence from the heart, is Low-energy anti-fibrillation pacing (LEAP). This method relies on the spatial heterogeneity in cardiac tissue to serve as nucleation sites for the production of intramural electrical waves that can target the sources of spatiotemporal chaos in cardiac tissue, thereby obliterating them~\cite{Luther2011,Chebbok2012}. 

The tendency of a spiral to break up in an excitable medium is critically determined by the medium's excitability. Spatiotemporal modulation of the excitability of the medium can change the dynamics of spiral waves.
Depending on the amplitude, frequency, degree of synchronization and the spatial scale of modulation of excitability, phenomena such as drift~\cite{PMID:22680565,Grill1996}, deformation~\cite{RAMOS20021383}, block~\cite{RAMOS20021383}, meander~\cite{Steinbock93,Grill1996}, breakup~\cite{RAMOS20021383,Schrader1995} and suppression of spiral waves~\cite{Ma2009} are known to occur.
In experiments, however, excitability is extremely difficult to control in real time and in a reversible manner. 
Thanks to optogenetics, this has recently become possible to a considerable extent.
Optogenetics has revolutionized cardiac research by successfully demonstrating supposedly pain-free real-time spatio-temporal control of spiral wave dynamics in cardiac tissue~\cite{Bruegmann2010,
Crocini2016,Bruegmann2016,Nyns2016,Bruegmann2018,Majumder2018}. With this tool one can regulate the flow of 
current across cell membranes to force these cell to trigger or suppress action potentials~\cite{Jia2011}. 
Although there is an extensive literature cataloguing the advantages and disadvantages of using optogenetics at supra-threshold light intensities (i.e. intensities capable of stimulating action potentials in individual cells or inducing waves in extended media), very little is known about the functionality and properties of the tool at the `sub-threshold' level. While some studies show that supra-threshold disturbances of spiral waves can cause wave breaks~\cite{Feola2017}, the corresponding response at sub-threshold intensities is not known.

An interesting study by Park et al.~\cite{Park2014} demonstrates the possibility to achieve prolongation of the action potential duration (APD) in optogenetically modified cardiomyocytes by applying optical stimuli in the subthreshold illumination regime. In particular, they show that optical pulses can prolong APD in a graded manner, thereby allowing one to ‘scuplt’ the action potential (AP) morphology in general. In another study by Burton et al.~\cite{Burton2015}, the authors demonstrate that application of a subthreshold light pulse can lead to an increase in conduction velocity of a propagating wave. However, this increase only occurs in the speed of the first wave in a wave train. The reason may be attributed to the current-voltage characteristics of the Channelrhodopsin current, which displays an initial spike, followed by saturation at a decreased level, when the optogenetically-mofidied cell is continuously exposed to light.

In this study we show a new mechanism of wave break initiation in simulated, optogenetically modified human atrial tissue in the presence of sub-threshold illumination. In particular, we observe that the global periodic pulsing of the domain with light of high sub-threshold intensity can lead to the development of wave breaks by "conditioning” of the wavelength, i.e. modulating the distribution of excitation along one arm of the mother spiral, thus creating an excitation window for wave break and re-entry.

\section{Results}

\begin{figure*}[t!]
\includegraphics[width=173.5mm]{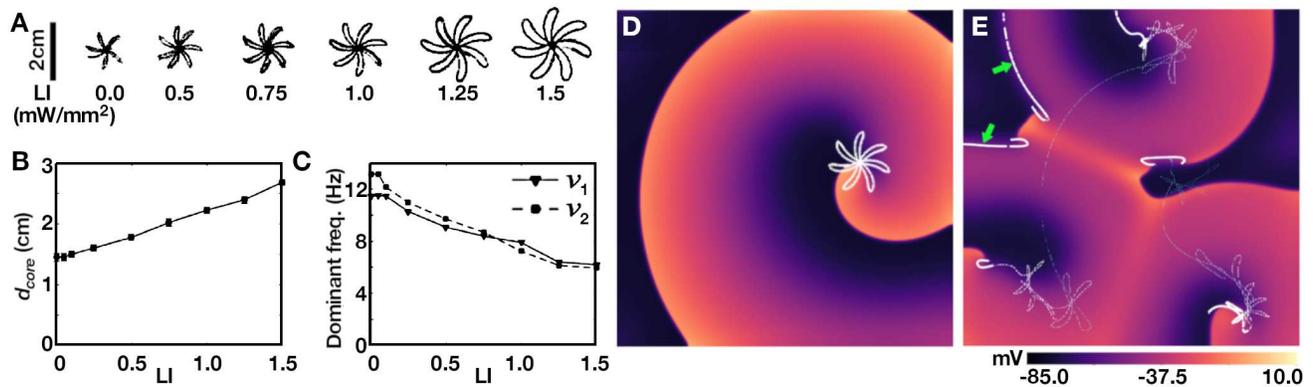}
\caption{Influence of a constant, uniform, global sub-threshold illumination with light intensity (LI) on spiral wave dynamics in 2D. A) Tip trajectory of the spiral at different LIs. B) Increase of the core diameter ($d_{core}$) of the spiral with increase of LI. Here the core is defined as the outer circle that just encloses the tip trajectory. C) Decrease in the dominant (first and second fundamental) frequencies of the spiral with increase in LI. D) Trajectory of the spiral wave at constant, uniform, global illumination below the threshold with LI = 0.75 mW/mm$^2$. E) Break up of the spiral wave at the same LI, when the light is applied periodically with a frequency of 2.1Hz. Green arrows point to the lines of the conduction block (shown in bold white lines). The faint grey lines trace the past dynamic history of the tips of the various spirals that have formed and terminated over time.}
\label{fig:characterization}
\end{figure*}

Under the influence of a constant, uniform global illumination, we found that an increase in light intensity (LI) led to an increase in the conduction velocity (CV) of the propagating wave in one dimension (1D) and to a progressive increase in the core size of the meandering spiral in two dimensions (2D), while maintaining the shape of the tip trajectory (Fig.~\ref{fig:characterization}A,B and D). The %number of lobes in the trajectory of the tip increased, while the 
dominant frequencies of rotation of the spiral decreased with increase in LI as illustrated in Fig.~\ref{fig:characterization}C. However, replacement of the time-independent illumination with a periodic perturbation led to the emergence of interesting dynamical effects. In particular,
\begin{figure}[b!]
\includegraphics[width=88mm]{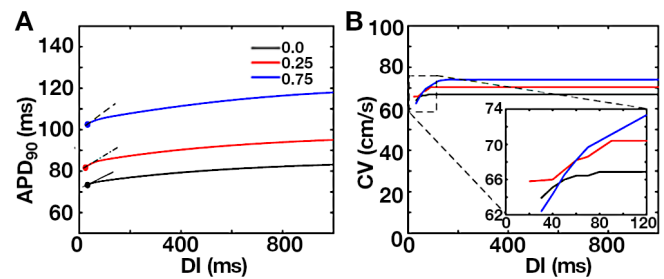}% Here is how to import EPS art
\caption{A) APD$_{90}$ and B) CV restitution curves in the absence of light (black), in the presence of uniform global illumination with low sub-threshold light intensity (LI=0.25mW/mm$^2$, red) and in the presence of uniform global illumination with high sub-threshold light intensity (LI=0.75mW/mm$^2$, blue). The slopes of the APD$_{90}$ restitution curves are 0.05, 0.06 and 0.16 for LI=0.0mW/mm$^2$ (black), LI=0.25mW/mm$^2$ (red) and LI=0.75mW/mm$^2$ (blue), respectively. The inset in (B) shows an enlarged part of the CV restitution curve at low diastolic interval (DI), which shows a crossover of the CV values.}
\label{fig:restitution}
\end{figure}
uniform, global, time-periodic light pulses of 0.75mW/mm$^2\le$ LI $<$ 2.0mW/mm$^2$ led to the initiation of wave breaks, as shown in Fig.~\ref{fig:characterization}E. These breaks appeared to occur right at the beginning of the `dark' phase of the applied stimulus, i.e., when the light was turned off in a pacing cycle. Application of high subthreshold LIs led to the spontaneous emergence of conduction blocks within the domain (see bold white lines indicated by green arrows in Fig.~\ref{fig:characterization}E), which seemed to promote wave break initiation. For the full sequence of events leading to the incidence of wave breaks, see Movie S1 of the SI Appendix. 

To explain this unusual observation, we first studied the influence of light on the dispersion of APD and CV, i.e. the restitution properties of the system. Our results, as shown in Fig.~\ref{fig:restitution}A and B, pointed to a strange anomaly: Both the APD and CV restitution curves for different LIs appeared to have gentle slopes (<1.0) without spatial dispersion, which is fascinating because this type of wave break intiation is typically associated with steep restitution curves (slope > 1.0), or dispersion in restitution properties within a system (contrary to our case), or in the presence of local ionic heterogeneities (as in case of ischemia)~\cite{Wu2002,Liu2004} or in domains with progressively degrading excitability, which results in the gradual flatenning of the CV restitution curve with reduced mean CV, characteristic of type-II ventricular fibrillation. %contradicts the general idea that wave breaks initiation is associated with steep restitution curves (slope > 1.0). 
In addition, constant application of light to the domain resulted in supernormal CVs, which was in direct conflict with our observation of the emergence of lines of propagation block, just prior to the occurrence of the break.

\begin{figure}[H]
\includegraphics{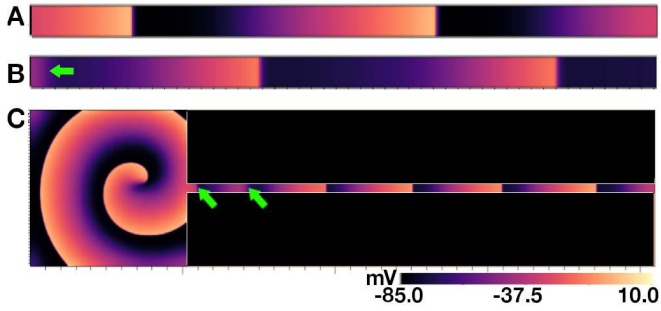}% Here is how to import EPS art
\caption{A) Right propagating plane waves in a pseudo-1D domain containing 512 $\times $ 10 grid points, electrically paced at 11.5 Hz from the left edge of the domain in the absence of illumination. Each stimulus results in the generation of a new propagating wave. B) Wave block at the pacing site when the simulation in (A) is performed in the presence of a uniform global time-periodic illumination, at 2.1 Hz temporal frequency and and LI=0.75 mW/mm$^2$. C) Wave block away from the site of pacing, in a pseudo-1D channel (1536 $\times $30 grid points) connected to a 2D domain (512 $\times $512 grid points) that maintains a spiral. The spiral drives the electrical stimulation in the channel. Time-periodic light stimulation is only applied to the pseudo-1D extension of the 2D domain. Green arrows indicate the locations of the wave block.}
\label{fig:pseudo1D}
\end{figure}

To understand the basis for this anomaly, we tried to reproduce the appearance of wave blocks in a reduced system, i.e. pseudo-1D, using a rectangular simulation domain containing 512$\times $10 grid points. We applied high frequency electrical stimulation to the left edge of the domain (Fig.~\ref{fig:pseudo1D}A), while illuminating the entire domain uniformly, in a time-periodic manner. In the absence of illumination, each applied stimulus resulted in the initiation of a new wave that propagated uninterruptedly to the right edge of the domain (Fig.~\ref{fig:pseudo1D}A). 
However, in the presence of an optical stimulation with LI$\ge$ 0.75 mW/mm$^2$ and any frequency between 2 - 12 Hz, we observed a random blocking of the stimulated waves, always at the site of stimulation ( Fig.~\ref{fig:pseudo1D}B). Thus, based on these data, it was impossible to conclude whether the block was actually a dynamic effect caused by optical stimulation or simply a result of CV restitution.

Given that wave block at the site of stimulation could give no clear insight into the formation of zero CV lines within the domain in 2D, we shifted the region for optical perturbation to 98 < x < 512, and resumed periodic electrical pacing at 11.5Hz (dominant frequency of the spiral wave in the absence of illumination), from the left end (x < 4) of the stripe. Our studies showed that every stimulus was capable of producing a new wave that propagated all the way to the right end of the stripe without interruption, although 1D traces along the length of the stripe did give us signatures of the effect of subthreshold stimulation on the membrane potential in the illuminated region. Thus, clearly, the 1D reduction of our 2D system was not optimised for wave block visualisation. To overcome this challenge, keeping in mind that the spiral in our 2D system was actually a meandering one, rotating quasi- periodically with 2 dominant frequencies, we chose to use the spiral in the 2D domain to drive the electrical activity in the stripe domain, while applying the optical perturbation only to the stripe domain. This had no direct bearing on the dynamics of the spiral wave itself, but was used as a driver of the electrical activity in the stripe domain, equivalent to the spiral in the 2D domain. Thus, we constructed a hybrid pseudo-1D equivalent of our original 2D system by connecting a 2048$\times $30 cell grid (pseudo-1D) to our 2D domain, which contained 512$\times $512 grid points. We initiated a spiral in the 2D domain. The waves emitted from the spiral core were able to move into the pseudo-1D lattice and thereby control the frequency of the applied electrical stimulation. We applied uniform global optical stimulation at LI=0.75mW/mm$^2$ and low frequencies ($\simeq$ 2.0HZ) only to the pseudo-1D domain, leaving the 2D domain unaffected by light. This setup allowed us to observe the formation of conduction blocks within the pseudo-1D domain, i.e. away from the stimulation site, over a wide range of frequencies of the applied light stimulation (Fig.~\ref{fig:pseudo1D}C, and Movie S2 of the SI Appendix).

\begin{figure}[b!]
\includegraphics{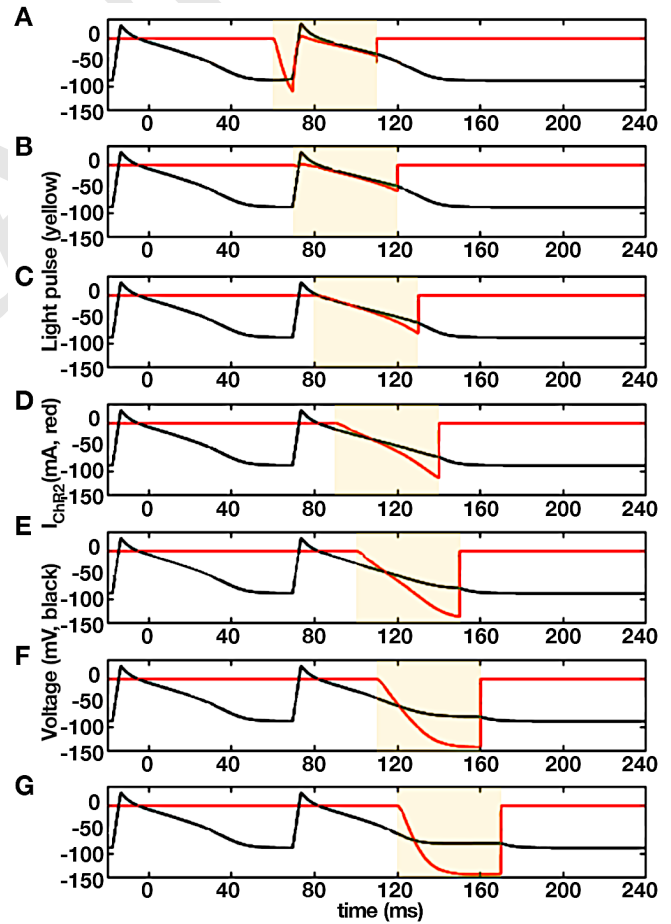}% Here is how to import EPS art
\caption{(A-G) Effect of applying a pulse of LI = 0.75 mW/mm$^2$ and duration 100 ms to different phases of an action potential (AP). For comparison, a control AP, i.e. in the absence of the light pulse, is provided next to the affected AP in each sub-figure. Black curves represent the transmembrane voltage (V), red curves show the corresponding channelrhodopsin current ($I_{ChR2}$), and yellow curves mark the duration of the application of the light pulse.}
\label{fig:single_cell}
\end{figure}

In addition to the disappearance of some of the waves that had propagated into the pseudo-1D domain over a certain distance, we observed spatio-temporal oscillation in the wavelength of successive waves.
 To understand the basis for these oscillations and to correlate to the appearance of the conduction block, we studied the effects of sub-threshold illumination on a single heart cell when exposed to light at different stages of an electrically stimulated AP (Fig.~\ref{fig:single_cell}). Our studies showed that for an optical stimulus that had approximately the duration of one AP, sub-threshold illumination had the greatest effect on the APD at $90$\% repolarization (i.e. APD$_{90}$), even when the light was applied during the recovery phase of the AP (Fig.~\ref{fig:single_cell}G).
\begin{figure}[b!]
\includegraphics{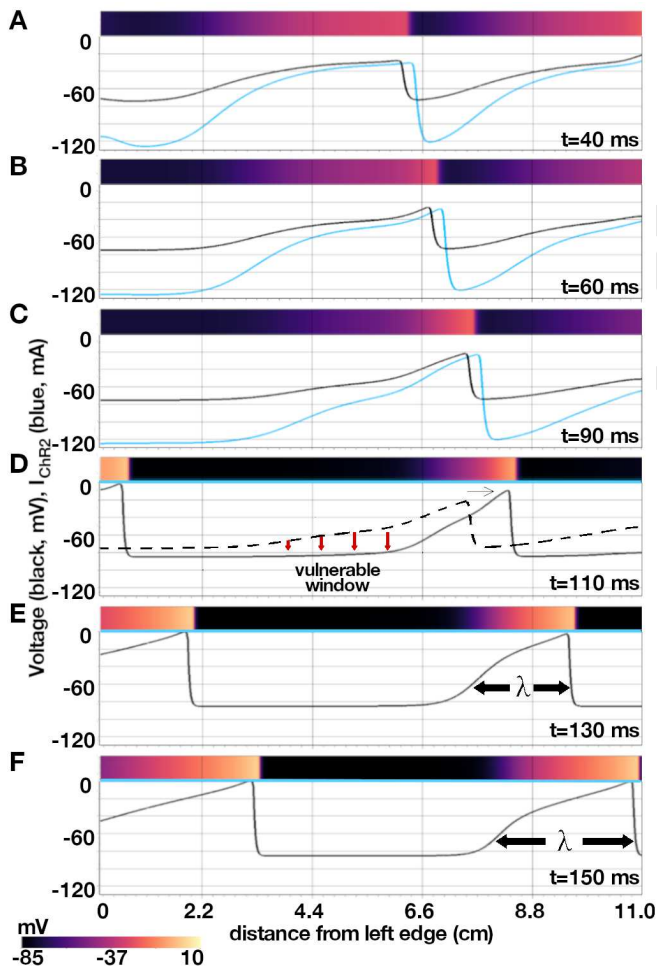}% Here is how to import EPS art
\caption{(A-C) Effect of applying a 100 ms, 0.75 mW/mm$^2$ light pulse to a plane wave propagating through a pseudo-1D domain containing 512 $\times$ 10 points. Black and blue curves represent $V$ and $I_{ChR2}$ respectively, measured from the line y=5 within the domain. (D) Rapid recovery of the wave tail (indicated by bold red arrows) immediately after the light is turned off. (E-F) Gradual recovery of the original waveform with increase in wavelength $\lambda$. In each sub-figure, the upper (colored) panel shows the spatial profile of the wave as it propagates through the pseudo-1D domain.}
\label{fig:vulnerable_window}
\end{figure}

A direct consequence of this differential response of the cell membrane to light applied at different phases of the AP is a phase-dependent delay of recovery, that is induced along the length of a wave, in extended media. Such a delay leads to differences in CV along a wavelength. Especially in the presence of light, the wavefront propagates at supernormal speeds while the waveback slows down. This results in (\textit{i}) prolongation of the wavelength with a gradual change in the shape of wave profile and (\textit{ii}) blocking of the following wave by the preceding one. When light is applied to a propagating wave, cells located at increasing distances from the wavefront towards the waveback, receive light at progressively shifted phases of their AP (Fig.~\ref{fig:single_cell}A to G).
This results in a gradual decrease in the rate of recovery of the cells in between the wavefront and the waveback, thereby changing spatial profile of the wave, to give it a pinched appearance. (Fig.~\ref{fig:vulnerable_window}A-C). Instead of the constant, uniform, global illumination, if the light is now applied periodically in time, the pinched wavelength gets the opportunity to return to its original state during the light-off period (Fig.~\ref{fig:vulnerable_window}D-F). This explains the occurrence of spatiotemporal oscillations in the wavelength of the propagating wave (in 1D) or a spiral arm (in 2D). 

\begin{figure}[b!]
\includegraphics[width=86.5mm]{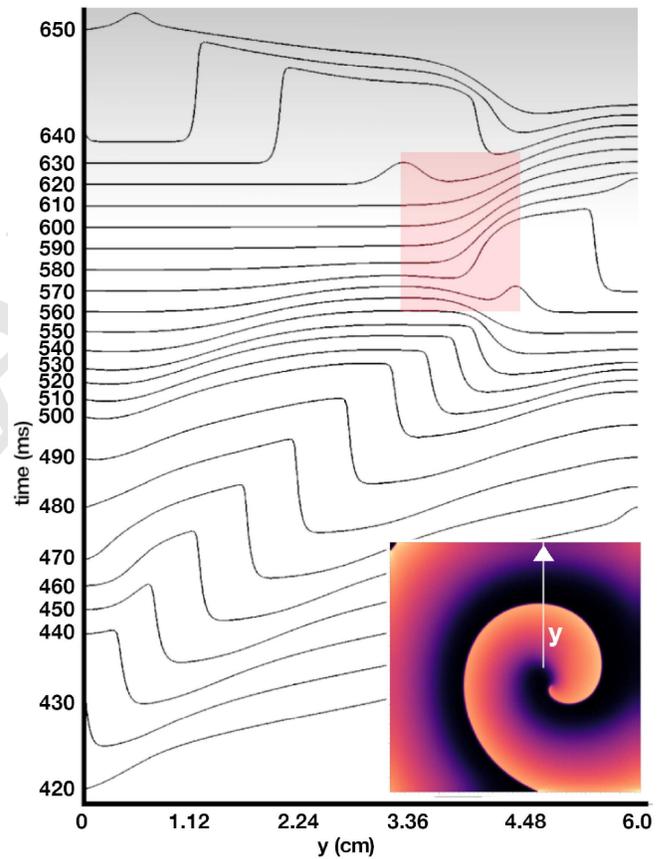}% Here is how to import EPS art
\caption{Demonstration of the mechanism of spiral wave breakup using a series of 1D traces of the spatial profile of the electrical activity in the medium, as measured along the solid white arrow shown in the inset. Here, y is the displacement from the center of the domain to the boundary, along the length of the arrow. The time stamp for each 1D trace is indicated on the left hand side of the vertical axis. The domain is illuminated uniformly and globally with LI = 0.75 mW/mm$^2$, from 430 ms to 610 ms. During this time, the CV of the right-propagating wavefront increases beyond normal, whereas, that of the waveback decreases. This manifests in a change in the spatial wave profile during the said period. Towards 570 ms, the wave profile modifies sufficiently to allow creation of a vulnerable window. Subsequent removal of light at 610 ms opens the window for re-excitation, which then permits re-entry; re-excitation occurs at 620 ms, which then gives rise to a left-propagating wave. The translucent pink rectangle marks the spatiotemporal window for wave break initiation. Reversal of wave direction is indicated on the figure using a colour gradient from grey, for left-propagating waves, to white, for propagation to the right.}
\label{fig:Mechanism2D}
\end{figure}

The higher the LI, the stronger the conditioning of the wavelength, i.e. the wave appears more pinched. The moment the light is switched off, the differences between the recovery rates of the cells at different distances from the wavefront disappear. The cells on the wavefront that allowed propagation at supernormal CV recover slowly, allowing the wavefront to expand from its pinched state in a gradual manner. The cells on the wave back that were forcibly held at a sub-threshold voltage higher than the normal resting membrane potential (RMP) in the pinched state recover immediately, creating a window for re-excitation. In 2D, this window allows re-entry, as shown in Fig.~\ref{fig:Mechanism2D}. The inset shows the spatial pattern of electrical activity in the domain, which assumes a spiral form. By tracing the voltage distribution along the solid white line (see Fig.~\ref{fig:Mechanism2D}, inset) in the direction of the arrow, we illustrate how a right-propagating wave is modulated over time with an optical perturbation (LI = 0.75 mW/mm2, from 430 ms to 610 ms) to create a vulnerable window (marked with a pink rectangle). The wave break occurs between 600ms and 620ms, leading to reexcitation and reentry, marked by the initiation of a left-propagating wave. Reversal of wave direction is indicated on the figure using a colour gradient from grey, for left-propagating waves, to white, for propagation to the right.

\begin{figure}[]
\includegraphics[width=86.5mm]{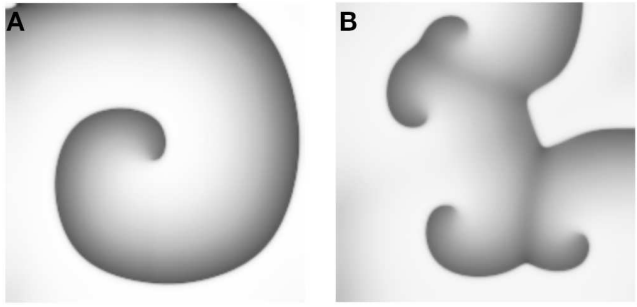}% Here is how to import EPS art
\caption{Spiral wave breakup in the 2D model for optogenetically-modified neonatal mouse ventricular tissue. A) Spiral wave in the presence of uniform, global, constant illumination at LI = 0.025mW/mm$^2$. B) Break up of spiral waves in the presence of periodic optical stimulation at 4 Hz, with a light pulse of duration 200 ms. %C) APD restitution curve and D) CV restitution curve at no illumination (black) and illumination at LI = 0.025mW/mm$^2$ (red).
}
\label{fig:wang-sobie}
\end{figure}

Finally, to test the generic nature of this new mechanism of wave break initiation at stable parameter r\'egimes, we repeated the study in another system, a 2D model of neonatal mouse ventricular tissue. At uniform global constant illumination with LI = 0.025 mW/mm$^2$, a spiral wave rotated with no signatures of breakup (Fig.~\ref{fig:wang-sobie}A). However, when stimulated periodically at 4 Hz, with light of the same intensity and pulse length 200 ms, breakup similar to Fig.~\ref{fig:characterization}E was observed (Fig.~\ref{fig:wang-sobie}B). A study of the restitution characteristics for this system, as presented in Ref.\cite{Mayer2017} (see Figures 5 A and B, therein), indicated that wave breaks occurred in a stable parameter r\'egime, thus proving the robustness and restitution-independence of our proposed mechanism for wave breaks.
%A deeper investigation into the restitution characteristics of the system reconfirmed that breakup indeed occurred in the parameter r\'egime for which the APD restitution curve had a gentle slope (??) and the CV restitution curve was relatively flat. These results are presented in Figure.~\ref{fig:wang-sobie}C-D.

\section{Discussion}
An intriguing question is: Would the choice of model parameters have an influence on wave break initiation, through a parameter-induced destabilizing effect (which may be characteristic of the cAF model used)? The answer is no. As such, the cAF model plays no specific role in determining the mechanism of wave break initiation and reentry. In Fig.~\ref{fig:wang-sobie} we demonstrate using a neonatal mouse ventricular model that such breaks can also occur in the ventricle. The reason for using the cAF model is because (\textit{i}) it makes more sense to ‘defibrillate’ a diseased condition rather than the healthy heart, where fibrillation is less likely to occur. We believe that any other parameter r\'egime that supports fibrillation (atrial or ventricular type) or tachycardia, should also demonstrate this phenomenon, under the application of pulsed subthreshold light stimulation. (\textit{ii}) The human atrial wall is super thin (~2-3mm) in contrast to the ventricles (~1.5cm), which makes the former a better candidate for optogenetic defibrillation, whose efficiency seems to rely on maximal illumination of bulk of the affected tissue. Researchers believe that the problem of poor penetration of visible light in cardiac tissue might not be so important in the human atria, because of its remarkably low wall thickness. Nevertheless, our studies reveal that one needs to be careful while designing the defibrillation protocol, as a series of globally illuminating light pulses in the suprathreshold regime, can, in fact trigger the formation of wave breaks underneath the surface, where illumination is subthreshold, and thereby lead to failure of defibrillation. Finally, (\textit{iii}) from a technical perspective, we chose the cAF model over the healthy heart model because of the large computational costs associated with the latter (see Methods for details). 

There are several theories regarding the factors and conditions that cause a spiral wave to break up. Of these, the strongest candidate is restitution (both APD and CV) in cardiac tissue~\cite{Fenton2002}. Some studies argue that APD and CV restitution curves fundamentally characterize the wave dynamics in the heart as they reflect the mesoscopic effects of changes in ion currents and concentrations occurring at the cellular level~\cite{Fenton2002}. 
Typically, a steep APD restitution curve (slope $\ge 1.0$) can drive the system via Hopf bifurcation to APD oscillations~\cite{Nolasco1968,Courtemanche1993,KARMA1994113}. Such alternans promotes the development of functional conduction blocks because the propagation of a "long" action potential (AP) into a "short" AP region fails, thereby limiting the allowable diastolic interval (DI)~\cite{Fenton2002}. Another proposed mechanism, again relating to the role of a steep APD restitution curve, suggests that the steepness of the curve is negatively correlated with 
the speed of the waveback in the presence of recovery gradients~\cite{Courtemanche1996}. A gradient $\ge 1.0$ introduces a difference between the CVs of the propagating wavefront and back so that within a wave train the front of the following wave collides with the back of the preceding wave, resulting in a functional conduction block~\cite{Courtemanche1993,Courtemanche1996a,Courtemanche1996,Watanabe2001}. Other mechanisms of wave break include the occurrence of a spatially non-synchronized type of APD oscillation, also known as spatially discordant APD alternans~\cite{Rosenbaum1994,Sato2006,Hayashi2007,Majumder2016a}, the presence of a biphasic APD restitution curve~\cite{Franz1988,Szigligeti1998}, effects of hysteresis~\cite{Lorente1990,Hall1999,Yehia1999,Oliver2000} and Doppler shift through the trajectories of the spiral peaks~\cite{Steinbock1992}. 

A fascinating study by Fenton et al.~\cite{Fenton99} on wave break initiation in ischemic heart tissue shows that dynamic flatenning of the APD restitution curve (characterized by slope < 1.0), results in an electrical ‘memory’ effect, which provokes a monotonic reduction in the excitability of cardiac tissue. Over time, the excitability becomes so low that a spiral breaks up close to its tip in the form of a `voltage drop’. The study concludes that spiral break-up can be induced in parameter r\'egimes with flat APD restitution as long as the dynamic effects of tissue memory and substantially large changes in basic cycle length are taken into account. Our study is consistent with the work of Fenton et al. in that wave breaks are observed at low APD and CV restitution slopes. However, we show that for a system with static APD and CV restitution curves, in the absence of a factor leading to a progressive reduction in tissue excitability and without explicit inclusion of the electrical memory effect, it is still possible to initiate wave breaks that develop into sustained re-entry. 

Another study by Zemlin et al.~\cite{Zemlin2001} reports spiral wave break-up in parameter regimes characterized by strongly negative slopes of the APD restitution curve. However, as noted in their work, such break-up is often less pronounced than the break-up that occurs with positive slopes, mainly because break-up with negative restitution takes a fairly long time to mature, compared to break-up with positive restitution. The wave break initiation that we demonstrate in this article happens very fast, in most cases, requiring not more than 1-2 pulses.

Another study worth mentioning in this context is that of Banville and Gray~\cite{Banville2002}, which demonstrates in experiments on the rabbit heart, the crucial role of the spatial distribution of APD- and CV restitution on the onset of alternans and fibrillation. However, in our study, we used a homogeneous domain with all cells identical, and subjected each cell to the same external conditions. Thus we did not have a dispersion of APD and CV restitution that could influence wave break initiation like in the case of~\cite{Banville2002}. Nevertheless, we observe the initiation of wave breaks, which leads us to rethink the underlying factors involved in initiating wave breaks in cardiac tissue under different conditions.

A rare, but quite efficient, mechanism of spiral wave break up involves supernormal excitability in certain domains. It is associated with a decrease in the excitation threshold with decreasing diastolic interval (DI). Typically, if it exists in a model, it promotes the development of conduction blocks through collisions between the rapidly advancing wave fronts and subsequent stacking at short DI, eventually leading to the initiation of wave breaks. Breakup occurs particularly close to the spiral tip. Supernormal CV can also lead to heterogeneity of refractoriness at shallow APD restitution, in which case scalloping occurs. The mechanism of spiral wave break initiation that we report in the present article is somewhat similar to this phenomenon. With uniform global light perturbations, we are able to impose supernormal CVs on the propagating electrical activity. This results in the development of wave blocks within the domain that are visible as white lines in Fig.~\ref{fig:characterization}E. In our studies, the reentry occurs once the factor causing supernormality is withdrawn, i.e., the light perturbation is removed, allowing the wavefront and wave back to restore their normal velocities.

Our study provides a clear example of a funny mechanism of wave break initiation and reentry that occurs when the slope of the APD restitution curve is less than 1.0 and the CV restitution curve is relatively flat. The proposed mechanism should be looked on as an important factor that can potentially lead to failure of optical defibrillation using pulsed global illumination. We believe that this mechanism is important to note because it provides useful insights into the translational prospects of optogenetic defibrillation in the human atria. And as our study proves, the phenomenon is non crucially dependent on the choice of the model used. Thus, we provides new insights into the possibility of the occurrence of wave breaks in heart tissue, even when there are no standard markers for vulnerability.  

Finally, the effect of global subthreshold stimulation was studied in simple ionic models of cardiac tissue, in the context of electrical defibrillation. A particularly close phenomenon was reported by Sridhar and Sinha~\cite{Sridhar2002}, who demonstrated the possibility to terminate electrical activity using a time-invariant global subthreshold electrical stimulation to achieve synchronization. Their study indicated that an electric pulse of length ~ $\mathcal{O}(1APD)$ was sufficient to cause this effect. Our studies showed (data not presented) that if the stimulus in\cite{Sridhar2002} was instead, applied at low frequency (~1-2Hz), using pulses of very short duration (~ $\mathcal{O}(0.1APD)$), then such perturbation could also initiate wavebreaks. However, such breaks occur within a very restricted parameter r\'egime and critically rely on the frequency of the applied perturbation for sustenance, owing to the sensitive dependence of the response of the single cell AP morphology, on the phase of the AP at which the stimulus is applied.

\matmethods{Electrical activity in cardiac tissue was modeled using the following reaction-diffusion-type equation:
\begin{equation}
    \frac{dV}{dt}=\nabla . \mathcal{D}\nabla V -\frac{I_{ion}}{C_m}
    \label{RDP}
\end{equation}
where $V$ represents the transmembrane potential (in mV) developed across single cardiomyocytes, $C_m$ is the specific capacitance (in ${\rm \mu F/cm^2}$) of the cell membrane, $\mathcal{D}$ represents the diffusion coefficient for intercellular coupling, and $I_{ion}$ represents the total ionic current produced by each individual cell. For human atrial tissue, $I_{ion}$ was formulated according to the Courtemanche-Ramirez-Nattel (CRN) model~\cite{Courtemanche1998}. 
It is a sum of $12$ ionic currents: fast $Na^+$ ($I_{Na}$), inward rectifier $K^+$ ($I_{K1}$), transient outward $K^+$ ($I_{to}$), ultra-rapid $K^+$ ($I_{Kur}$), rapid and slow delayed rectifier $K^+$ ($I_{Kr}$ and $I_{Ks}$, respectively), $Na^+$ and $Ca^{2+}$ background ($I_{BNa}$, and $I_{CaL}$), $Na^+/K^+$ pump, $Ca^{2+}$ pump ($I_{pCa}$), $Na^+/Ca^{2+}$ exchanger ($I_{NaCa}$) and $L-type~Ca^{2+}$ current ($I_{CaL}$). 
The parameters of the CRN model were adjusted to reproduce the action potential of the normal atrial working myocardium~\cite{Tobon2013}. In particular, the maximum conductance for $I_{K1}$, i.e., $G_{K1}$ was changed from 0.09 nS/pF in the original model, to 0.117 nS/pF, to yield a single cell action potential duration at 90\% repolarization of the membrane potential (APD$_{90}$), of 284 ms. 
A choice of $\mathcal{D}$=0.0023 cm$^2$/ms, produced a conduction velocity (CV) of 69.75 cm/s in the two-dimensional healthy tissue domain. In order to model the action potential during chronic atrial fibrillation (AF) remodelling, 
%we used a combination of parameters sets from earlier publications~\citep{Courtemanche1999,PANDIT20053806,Zhang2005}. In particular, 
the maximal conductances of $I_{to}$, and $I_{CaL}$ were reduced by $85\%$, and $74\%$, respectively, $G_{K1}$ was increased by $250\%$, the time constant for activation of $I_{CaL}$ was increased by $62\%$, the activation curves for $I_{to}$ and $I_{CaL}$ wer shifted by +16 mV and -5.4 mV, respectively, while the inactivation curve for $I_{Na}$ was shifted by +1.6 mV~\cite{Courtemanche1999,Zhang2005,PANDIT20053806}. These parameter adjustments reduced the wavelength (at 1Hz electrical pacing) from $\simeq$18 cm in healthy tissue, to $\simeq$5 cm, allowing us to fit the spiral into a smaller simulation domain (512 $\times$ 512, as opposed to 2048 $\times$ 2048) without causing it to break up. This improved the computational cost-effectiveness by a factor of 20. The spiral wave in our study meandered with a hypocycloidal tip trajectory, and survived for longer than 10s of simulation time. 

For the neonatal mouse ventricular model, $I_{ion}$ was expressed as a sum of 16 ionic currents, according to the Wang and Sobie~\cite{wangsobie.2007}. These include, the fast $Na^+$ current ($I_{Na}$), the background $Na^+$ and $Ca^{2+}$ currents
($I_{Nab}$ and $I_{Cab}$), the L-type and T-type $Ca^{2+}$ currents  ($I_{CaL}$ and $I_{CaT}$), the $Ca^{2+}$ pump current ($I_{pCa}$), the $Na^+/Ca^{2+}$ exchanger ($I_{NaCa}$), $Na^+/K^+$ pump ($I_{NaK}$), slow and
fast components of the transient outward $K^+$ currents ($I_{Kto,f}$ and $I_{Kto,s}$), slow and rapid delayed rectifier $K^+$ currents ($I_{Ks}$ and $I_{Kr}$), ultrarapid delayed rectifier $K^+$ current ($I_{Kur}$), sustained outward $K^+$ current ($I_{Kss}$), inward rectifier ($I_{K1}$), and the
$Ca^{2+}$-activated $Cl^-$ current ($I_{Cl,Ca}$). We used $\mathcal{D}$=0.00095 cm/ms, which led to a CV of 43.9 cm/s. The 2D simnulation domain contained 200 $\times$ 200 grid points.

In order to incorporate the effects of optogenetics, our models were combined with a 4-state model for voltage- and light-sensitive Channelrhodopsin-2~\cite{Williams2013}, as described in Eqs.~\ref{Optomodel1}-\ref{Optomodel5}.
\begin{eqnarray}
I_{ChR2} = g_{ChR2} G(V) (O_1 + \gamma O_2) (V-E_{ChR2},  \label{Optomodel1}\\
\frac{dC_1}{dt} = G_r C_2 + G_{d1} O_1 - k_1 C_1,\\
\frac{dC_2}{dt} = G_{d2}O_2 - (k_2 + G_r) C_2,\\
\frac{dO_1}{dt} = k_1C_1 - (G_{d1} + e_{12}) O_1 + e_{21} O_2,\\
\frac{dO_2}{dt} = k_2C_2 - (G_{d2} + e_{21}) O_2 + e_{12} O_1,\\
O_1+ O_2 + C_1 + C_2 = 1
\label{Optomodel5}
\end{eqnarray}
%This model describes the kinetics of an ion channel that is added to the existing repertoire of ion channels in the cell. 
Here $O_1$, $O_2$, $C_1$ and $C_2$ represent the open and closed states of the ChR2 ion channel. $k_1$, $k_2$, $G_{d1}$, $G_{d2}$, $G_r$, $e_{12}$, and $e_{21}$ are the kinetic parameters, whereas, $G_r$, $G(V)$ and $G_{d1}$ contain the voltage-dependencies. For a detailed description of these parameters and their values, we refer the reader to~\cite{Williams2013}.
The total current produced by each cell in Eq.~\ref{RDP} is $I_{ion}$, plus, the Channelrhodopsin current $I_{ChR2}$. We found that for human atrial tissue, LI < 2.0mW/mm$^2$ were typically sub-threshold, meaning, they failed to trigger action potentials, whereas, higher light intensities successfully stimulated cells and initiated waves in extended media. In neonatal mouse, LI < 0.03 mW/mm$^2$ was considered sub-threshold.
}

\showmatmethods{} % Display the Materials and Methods section

\acknow{This work was supported by the Max Planck Society and the German Center for Cardiovascular Research (DZHK).}
\showacknow{} % Display the acknowledgments section

\section*{SI Appendix}
\textbf{Movie S1:} Controlled break-up of spiral waves, induced by periodic sub-threshold optical stimulation, at frequency 1.56 Hz. The trajectory of the spiral tip(s) is traced using white lines that fade over time. The filled blue circle below the voltage color-code (in mV) in select frames indicates the application of uniform, global subthreshold illumination at LI = 0.75 mW/mm$^2$, in these frames.\\
\\
\textbf{Movie S2:} Reduction of the 2D system to a hybrid 1D system, in which we combine a 2D simulation domain to a pseudo-1D outlet. The electrical stimulation in the pseudo-1D domain is driven by the unperturbed spiral wave in 2D. A uniform global periodic sub-threshold stimulation (frequency: 2.1 Hz) with light of intensity 0.75 mW/mm$^2$ is applied to the pseudo-1D domain. Propagation failure occurs well within the pseudo-1D domain for randomly selected waves. In addition, we observe oscillations in wavelength and modulation of the wave profile for waves propagating through the pseudo-1D domain. 

% Bibliography
\bibliography{pnas-sample}

\end{document}